\documentclass[12pt]{article}
\begin{document}

\begin{center}
{{\bf Gauge/String-Gravity Duality and Froissart Bound\footnote{\uppercase{P}resented to the  \uppercase{C}oral \uppercase{G}ables \uppercase{C}onference 2003, ``\uppercase{L}aunching of \uppercase{B}elle\uppercase{E}\'poque in \uppercase{H}igh \uppercase{E}nergy \uppercase{P}hysics and \uppercase{C}osmology", 17 - 21 \uppercase{D}ecember, 2003,    \uppercase{F}ort \uppercase{L}auderdale, \uppercase{F}lorida.}} }
\end{center}

\begin{center}
{\bf KYUNGSIK KANG} 
%\normalsize
\\
 Department of Physics, Brown University, \\
Providence, RI 02912, U.S.A. \\
{\it e-mail: kang@het.brown.edu}
\end{center}

\begin{abstract}
The gauge/string-gravity duality correspondence opened renewed hope and possibility to address some of the fundamental and non-perturbative QCD problems of in particle physics, such as hadron spectrum and Regge behavior of the scattering amplitude at high energies.  One of the most fundamental and long-standing problems is the high energy behavior of the total cross-sections.  
According to  a series of exhaustive
tests by the COMPETE group, (1) total cross sections have a universal Heisenberg behavior in energy corresponding to the maximal energy behavior allowed by the Froissart bound, i.e., $ A + B ~~ \ln ^{2} (s/s_0) $ with $ B \sim 0.32 ~~ mb$ and $s_0 \sim  34.41 ~~ GeV^2$ for all reactions, and (2). the factorization relation
among $\sigma _{pp, even}, \sigma _{\gamma p}$ and $\sigma _{\gamma \gamma}$ is well satisfied by experiments. I discuss the recent interesting application of the gauge/string-gravity duality of $AdS / CFT$ correspondence with a deformed background metric so as to  break the conformal symmetry that  lead to the Heissenberg behavior of rising  total cross sections, and present some preliminary results on the high energy QCD from Planckian scattering in AdS and black-hole production.
\end{abstract}

%\end{center}

\section{ Global Description of High Energy Scattering - COMPETE results}

The COMPETE group\cite{one} has performed a series of exhaustive tests of the analytic parametrisations for the forward scattering amplitudes against the largest available date at $t=0$, which includes all measured total cross sections as well as the ratios, $\rho$, of the real part to imaginary part of the elastic amplitude of $pp , \bar{p}p, \pi^{\pm} p, K^{\pm} p$, and total cross sections of $\gamma p, \gamma\gamma$ and $E^- p$.  Applying a set of carefully designed criteria for measuring the quality of fits to differentiate the different parametrizations, beyond the usual $\chi^2/dof$, they found $RRPL2_u$ and $RRP_{nf}L2_u$ to be the best analytic amplitude models.  Common features of these models are (1) total cross sections have a universal Heisenberg behavior in energy corresponding to the maximal energy behavior allowed by the Froissart bound.  i.e. \\
 $B~\log^2~(s/s_0)$ for all total cross sections with\\
$B = 0.315 mb, s_{0} = 34.41 GeV^2$ for $RRPL2_{u}$  and\\
$B = 0.315 mb, s_{0} = 34.03 GeV^2$ for $RRP_{nf}L2_{u}$  irrespectively of degeneracy in Reggeon terms, though favoring non-degeneracy:  for example, 
$B= 0.328mb, s_0 = 49.06 GeV^2 $ for $ (RR)_d PL2_u$.

\noindent (2)  Total cross sections satisfy the
 {\bf factorization relation}, $(H_{\gamma p})^2 \rightarrow H_{\gamma \gamma} \times H_{p p}$,  by the  $H = PL2$ terms. Numerically, $\delta = (H_{\gamma p}/ H_{pp}) = 0.0031$, \\
in good agreement with the generalized vector dominance.  

\section{Theoretical Models for Rising Cross Sections}

Assuming that the hadron-hadron scattering at high energies as a collision of two flat discs that produce and exchange a pair of mesons in the interaction region of the impact parameter space and that the portion of the energy density that is responsible for the non-renormalizable meson exchange interactions is high enough to create at least a pair of mesons and the portion is exponentially decreasing with the exchanged meson mass and impact distance in analogy to the shock wave process,  Heisenberg\cite{two} has argued that the maximum impact distance for which the effective interaction takes place, corresponding to the minimum portion to create a pair of mesons, is $b_{max} = (1/2m) \log (s/s_0 ), m$ being the exchanged meson pair mass.  From this the total cross section is given by
$$
   \sigma = ~ (\pi /16 m_{\pi}^2) ~ log^2 ~ (s/s_0) 
$$  
which corresponds to a saturating behavior of the {\bf Froissart bound}\cite{three}
$$ \sigma_{tot} \leq c \log ^2 s, 
$$  in which 
$$c \leq (\pi /m_{\pi}^2) = 60 mb$$
which is a consequence of the unitarity and positivity of the imaginary part of the scattering amplitudes in the Lehmann ellipse.

The increasing behavior of the total cross sections is a necessary condition for the rigorous proof of the Pomeranchuk theorem\cite{four}.  Also one can show from unitarity and analyticity in the form of the derivative dispersion relation that the $s-u$ crossing symmetric, the forward scattering amplitude that saturates the Froissart bound, is of the form $PL2$ from\cite{five}.

\section{ Gauge/Gravity Duality and the Heisenberg Behavior}

The gauge/string-gravity duality of the $AdS_{d+1}/CFT_d$ correspondence\cite{six}, i.e., the weak coupling gravity of superstring theories in $AdS$ space of $d+1$ dimension is dual to the strong coupling conformal supersymmetric gauge theory in $d$ dimension, has opened the possibility to address the high energy behavior of the scattering amplitude, in particular, the high energy behavior of the total cross sections.  But in order to deal with a realistic QCD relevant for the hadron physics, the desired super-gravity solutions must embody the salient features of strong interactions, such as confinement, hadronization with non-zero mass gap and Regge behavior of the hadron scattering amplitude at high energies.  On the other hand, the super-string theory contains no mass gap due to zero mass gauge/graviton fields in the string spectra of asymptotic states, and has soft scattering amplitude,
$$
 A_{string}(s, t) \sim \exp [ {-\alpha'\over 2} (s \ln s + t \ln t + u \ln u) 
]
$$
to be contrasted to the partonic or hard Regge behavior in gauge theory,\\
$$ A_{qcd} (s, t) \sim s^{2 - {n \over 2}}
$$
Can one find a consistent picture of gauge field properties and Regge amplitude in the strong gauge coupling regime from a suitable string theory via $AdS/CFT$ correspondence?

A suggestion\cite{seven} is to deform the Randall-Sundrum type AdS/gravity background metric with an IR cutoff in the holographic radial coordinate $r$ of a space and in particular use a metric of the form,
$$
ds^2 = (r/R)^2 {\eta _{\mu \nu}} dx^{\mu} dx^{\nu} +  (R/r)^2 (1 - (b/r)^{d} )^{-1} dr^2 + R^2 ds_{Y} ^2 
$$
\noindent
whereby breaking the conformal (and SUSY) symmetry with an IR cut-off at $r_{min} = b$.  Here $R$ is the anti-deSitter radius and $ds_{Y} ^2$ is the metric for 5 (or 6) compact dimensions of 10-d string (or 11-d M-) theory.

One of the most interesting features of this background metric is the warp factor multiplying the $4-d$ flat metric which leads to the holographic relation between $r$ and $4-d$ momentum $p$. 
Basically in 10-d string theory, due to the gravitational {\bf Red Shift} for a state localized in the transverse  co-ordinate $r$, $ \Delta s = (R/r) \Delta r $ and $p = (r/R) p_{s}$, so that a state with a characteristic 10-d energy scale, $p_{s} \sim (1/R)$, corresponds to a 4-d energy scale $p \sim (r/R^{2})$,  resulting  the holographic relation $\colon$ high energy is large r and low energy is small $ r$.
Depending on the structure of the extra $5-d$ or $6-d$ compact space and the details of the geometry at small $r$, the precise gauge theory and the breaking of  the conformal invariance are different but the high energy QCD is well approximated by this background metric independently of the details of the extra dimensional space at low $r$ near the $IR$ cutoff point 
 $r_{min} \sim \Lambda_{KK} R^{2}, \Lambda_{KK}$ being the conformal scale, i.e., the mass gap determined by the lightest glue ball ($KK$ mode).  In other words, such theory embodies the QCD universalities in the sense that it is approximately conformal in the large momentum region while it has a non-zero mass gap and confinement in the infrared region.  

Polchinski and Strassler\cite{eight} have argued that with the extra dimensional branes of either warped or non-warped large space-time geometry, the amplitude can be treated essentially as $10-d$ scattering that takes place at a point in $AdS$ in which transverse dimensions are integrated coherently over, so that the soft behavior of the strings would conspire the shape of the bulk wave functions and produce the correct power behavior of the confining gauge theory.  In what follows, I will present the analysis of the gauge/gravity following the papers under preparation with H. Nastase\cite{nine}. 

 A glueball corresponds to a plane wave state $\psi (r, Y) e^{i x_{\mu} p^{\mu}}$  in $AdS$ and scatters with a local proper momentum $p_s (r) 
= (R/r)p$, i.e., UV shifted in the IR.  There is a gauge theory string tension
\begin{equation}
\alpha^{\prime} = (gN)^{-1/2} \Lambda_{KK}^{-2}
\end{equation}
and
$$
\sqrt{\hat{\alpha}^{\prime}} p_{s} = \sqrt{\hat{\alpha}^{\prime}} p  (r_{min}/r ) \leq \sqrt{\hat{\alpha}^{\prime}} p
$$
Note that at small $r, p_s$ is larger than the string scale
 $ 1/\sqrt{\hat{\alpha}^{\prime}}$ and $A_{string}$ has the soft behavior.

Gauge theory scattering of glueballs is equated to a scattering inside $AdS$ of the above states via $AdS/CFT$ correspondence, leading to 

\begin{equation}
\mathcal{A}  (p) = \int \, drd^5 \Omega \sqrt{g} A_{string} (\vec{p} ) \prod_i  \psi_i
\end{equation}

The integral is damped by the soft behavior of $A_{string}$ at small $r$ and also by the wave functions at large $r$, since $\sqrt{g}=r^3 R^2$, with $\psi \sim Cf(r/r_{min})g (\Omega) \sim C(r/r_{min})^{-\Delta}g(\Omega)$.  We get in the scattering region,  $\sqrt{\hat{\alpha}^{\prime}} p  \gg 1, r_{scatt} \gg r_{min}$ and
\begin{equation}
\mathcal{A}  (p) \sim \int \, drr^3 \prod_i  \left({r_{min}\over r}\right)^{\Delta_{i}} \, \mathcal{A}_{string} (pR/r) \sim \left( {\Lambda_{KK}\over p} \right)^{\Sigma\Delta_i-4}
\end{equation}

as in QCD. Note this is just from scaling of the amplitude in the scattering region $r_{scatt} \gg r_{min}$ where $10-d$ scattering amplitude for $2\rightarrow 2$ is dimensionally $g^2 (\hat{\alpha} )^3 F (p_s \sqrt{\hat{\alpha}^{\prime}} ).$  
\noindent
In the Regge limit $0< -t \ll s$, with $\hat{s} \gg |\hat{t}|, \hat{\alpha}'^{-1}$,  
\begin{equation}
\mathcal{A}_{string} (p_x) \sim F \left(p_s \sqrt{\alpha^{\prime} }\right) \sim (\hat{\alpha}^{\prime} \hat{s} )^{\alpha ' t/2+2} \, {\Gamma (-\hat{\alpha}\hat{t}/4)\over\Gamma (1 + \hat{\alpha}^{\prime} \hat{t}/4)}
\end{equation}
from the Virasoro-Shapiro amplitude for massless external states $(s + t + u = 0)$. With this $\mathcal{A}_{string}$, we have, using $\nu = - \hat{\alpha}^{\prime} \hat{t} $ as integral variable,

\begin{equation}
\mathcal{A} \sim {1\over (\alpha ' \vert t \vert )^{\Delta/2-3} } \int_0^{\nu_{max}} d \nu\nu^{\Delta /2-3} \, A_{string} (\nu s /\vert t\vert , \nu )
\end{equation}
since $s/t = \hat{s}/\hat{t}$, and $r = R (\alpha^{\prime} t/\nu)^{1/2}$.

The Regge behaviour can then be obtained if the dominant contribution of the integral comes from the upper limit, $\nu_{max}$(which comes from $r_{min}$), which is $\alpha^{\prime} \vert t\vert $,   because the saddle point of the integrand is outside of $\nu_{max}$, 

$$\alpha'|t|< (\Delta -4)/ln(s/|t|),$$ 
so that the flat space Regge behavior $A \sim (\alpha ' s)^{2+\alpha^{\prime} t/2}$ follows apart from some $t$ factor due to different metrics and wave-function in $r$.  Otherwise, we get
\begin{equation}
\mathcal{A} \mathnormal \sim s^2|t|^{-\Delta/2}[ln(s/|t|)]^{1-\Delta/2}
\end{equation}
and the main contribution to the integral comes from again far away from the cutoff, i.e., $r_{scatt} \gg r_{min}$. Note the inverse power of $\vert t \vert$ in (6) for any $t$ and sufficiently large $s$.

Subsequently {\bf Giddings}\cite{ten} studied other bulk perspectives and in particular argued that the effect of strong-gravity processes, such as black hole formation, to the high energy behavior of the total cross sections is important in the dual dynamics. Here the key point is that in TeV scale gravity scenario, black holes should be produced once the energy passes the fundamental Planck scale near a few TeV.   

Giddings points out that as one increases the energy of the gauge theory scattering, one increases also the relevant energy in string theory. And there are three further (higher) scales (in the case when the string coupling $g_s$ is small but $g_sN$ is large). 

The first is the Planck scale 
$$M_P \sim g_s^{-1/4}/\sqrt{\hat{\alpha'}}, $$
 which corresponds in the gauge theory to $\hat{M}_P = g_s^{-1/4}/\sqrt{\alpha'}$, where black holes start to form. Note $\hat{\alpha'} = \Lambda_{KK}^{-2}(g_sN)^{-1/2} = \alpha '$ and $r_{min}=R=1/\Lambda_{KK}$. The black hole production cross section is approximated  by $\sigma \simeq \pi r_H^2 \sim E^{2/7}$ in $10-d$ flat dimension, since we have approximately $10-d$ flat space.

\noindent
The second scale is where string intermediate states cross over to black hole virtual states,
$$
E_c \sim g_s^{-2}/\sqrt{\alpha'} = N^{7/4}M_P/(g_sN)^{7/4} = N^2 \sqrt{\frac{\Lambda_{KK}}{r_{min}}} \frac{1}{(g_sN)^{7/4}}
$$\\
or in the gauge theory
$$
\hat{E}_c \sim \frac{N^2 \Lambda_{KK}}{(g_sN)^{7/4}}
$$

Here the semi-classical result should be also applicable to the cross section.\\

The third scale is when the black hole size $r_H $ is comparable to the AdS size R, 
$$
E \sim M_P^8 r_H^7 \left(M_P^{d-2}r_H^{d-3} {\rm \,\, in\,\, general}\right) \rightarrow E_R = M_P^8R^7
$$\\
or in the gauge theory $
\hat{E}_R = N^2\Lambda_{KK} $ At this energy, $r_H \sim lnE$ so that $\sigma \sim ln^2E$ corresponding to the maximal Froissart bound.

Though no proper dynamical theory of black-hole production was used, one may say that Giddings argument may represent some features of scattering in tran-Planckian energy regime.  Here two particle scattering at small impact parameter region would induce a large space-time curvature, larger than the fundamental scale, due to a large concentration of energy in a small volume.  To a low energy observer, this will appear as a curvature singularity whose horizon will hide the high energy effects from the observer outside.  Also it has been known\cite{eleven} that trapped surfaces do form in the classical trans-Planckian collision of particles.  Trapped surface will cause a space-time singularity\cite{twelve} which will give the horizon censorship.  Apparent horizon area can not be smaller than that of the event horizon, which in turn can not be smaller than the black-hole area.

Giddings approximates warped metric by an AdS type
$$
ds^2 \approx \frac{R^2}{z^2}(dz^2 + \eta_{\mu\nu} dx^{\mu}dx^{\nu}) + R^2ds^2_Y
$$
by changing the variable as $z=R^2/r=Re^{y/R}$. Then  $z \in (0,R=r_{min})$ corresponds to $y \in (-\infty,0)$. In particular, the gauge theory brane is located at the IR cut-off $z=R=r_{min}$. Notice that the confining gauge theory dual is the same as the Randall-Sundrum\cite{thirteen} scenario 2 with the gauge theory living in the $IR$ brane where the warp factor is minimal and with the Planck brane at $z=0$.

The perturbed $AdS$ metric is 
$$ ds^2 = (1 + h_{yy}) dy^2 + \exp (-2y/R) (\eta_{\mu\nu} + h_{\mu\nu} ) dx^{\mu} dx^{\nu} + R^2 ds_Y^2
$$ 
with the gauge $h_{\mu y} = 0$. In particular, Giddings assumed a black-hole solution due to a static point mass source living in the $IR$ brane and extending to the bulk by putting
$$
T_{\mu\nu}=S_{\mu\nu}\delta(y), T_{yy}=T_{y\mu}-0
$$
where $S_{\mu\nu}$ is a static point mass source
$$
S_{\mu\nu}=2m\delta^{d-1}(x)\delta_{\mu 0}\delta_{\nu 0}
$$
 in the linearized gravitational equations, which he solved by using Neumann's Greens function.  The most interesting component of his solution is
\begin{equation}
h_{00} \simeq \frac{k}{RM_P^{d-1}}m(\frac{z}{R})^d\frac{e^{-M_1r}}{r^{d-3}}
\end{equation}
either in light or heavy radion region $\mu R \ll 1$ or $\mu R \gg 1$ where $M_1$ is the mass of the lightest KK mode of graviton, the first pole in the integral expression for $h_{00}$ coming from the first zero of the Bessel function $J_{d/2 -1} (qR)$, $q^2$ being $-p^2$.  However for $\mu R \ll 1$ the linear approximation for neglecting brane bending (due to the point mass) is lost before reaching the horizon. So $r_H$ cannot be obtained from this solution even approximately.\\
\noindent
But in the heavy radion limit $\mu R \gg 1$, the solution gives an estimate of the horizon size $r_H=r_H(m,y)$ from $h_{00} \sim -1$.  
In particular, at the IR brane, corresponding to $y=0$, the black hole size is given by 
$$
r_H(m,y=0) \sim \frac{1}{M_1}ln \mid \frac{kmM_1^{d-3}}{RM_P^{d-1}}\mid \sim {1\over M_{1}} ln m
$$
so that 
$$
r_H(E) \sim ln E
$$ 
and 
$$ \sigma \sim \frac{1}{M_1^2}ln^2 E
$$
In the intermediate case, if $M_1 < M_L$, the horizon forms first, while if $M_1 > M_L$, brane bends first. But even if $M_L$ is smaller, the radion $L$ has the same expression or more precisely $LM_L$ is $h_{00}$ with $M_1$ and $1/R$ replaced by $M_L$, which means 
$$
\sigma \sim \frac{\pi}{M_L^2} ln^2 E
$$
The essential point of getting the Froissart-Heisenberg behavior for the cross section is the appearance of the exponential term with the non-zero mass gap.  But we find that\cite{nine} there is no such exponential in the $AdS$ Aichelburg-Sexl waves\cite{thirteen} for $r\gg R$.  In fact it gives just the regular $4-d$ behavior in this limit in Randall-Sundrum\cite{fourteen} for string-like models of Giddings et al\cite{fifteen}, prompting us\cite{nine} to examine critically how $h_{00}$ is obtained under $ r \gg R $ and/or $\exp (r/R) \ll 1$, as well as how the trapped surface in $4-d$ was derived by Eardley and Giddings\cite{sixteen}.

\noindent{\bf Acknowledgment}

I would like to thank H. Nastase for the on-going collaborations and enlightening discussions.  This talk is the improved version of the earlier presentations which  can be found in part in Ref. [17].  This work is supported in part by the U.S. D.o.E. Contract DE-FG-02-91ER40688-Task A, with the Report number Brown-HET-1403,


\begin{thebibliography}{99}
\bibitem{one}J. R. Cudell et al, {\it Phys. Rev. Lett.} {\bf 89}, 201801 (2002), hep-ph/0206172;  
J. R. Cudell et al, {\it Phys. Rev.} {\bf D65}, 07024 (2002), hep-ph/
0107219; 
See also the COMPETE webpage http://nuclth02.phys.ulg.be/compete.

\bibitem{two} W. Heisenberg, {\it Z. Phys.} {\bf 133} (1952) 65; See also H. G. Dosch, P. Gauron and B. Nicolescu, {\it Phys. Rev.} {\bf D67} (2003) 077501.

\bibitem{three} M. Froissart, {\it Phys. Rev.} {\bf 23} (1961) 1053; L. Lukaszuk and A. Martin, {\it Nuovo Cimento} {\bf A52} (1967) 122.
\bibitem{four} Y. Ya. Pomeranchuk, {\it JETP} {\bf 7} (1958) 499.

\bibitem{five}  K. Kang and B. Nicolescu, {\it Phys. Rev.} {\bf D11} (1975) 2461.

\bibitem{six}  J. Maldacena, {\it Adv. Theor. Math. Phys.} {\bf 2} (1998) 231; E. Witten, ibid 253; S. S. Gubser, I. R. Klevanov and A. M. Polyakov, {\it Phys. Lett.} {\bf B428} (1998) 105.

\bibitem{seven}  E. Witten, {\it Adv. Theor. Math. Phys.} {\bf 2} (1998) 505.

\bibitem{eight}  J. Polchinski and M. J. Strassler, {\it Phys. Rev. Lett.} {\bf 88} (2002) 03160.

\bibitem{nine}  K. Kang and H. Nastase, ``High Energy QCD from Planckian Scattering in AdS" (in preparation), and ``Planckian Scattering Effects ad Black Hole Production in Low $M_{pl}$ Scenarior", (in preparation).

\bibitem{ten}  S. B. Giddings, {\it Phys. Rev.} {\bf D67} (2003) 126001, hep-th/0203004.

\bibitem{eleven}  P. D. D'Eath and P. N. Payne, {\it Phys. Rev.} {\bf D46} (1992) 658; 675; 694.

\bibitem{twelve}  S. W. Hawkings and G. F. R. Ellis, ``The Large Scale Structure of Space-Time", Cambridge University Press, 1973,

\bibitem{thirteen}  L. Randall and K. Sundrum, {\it Phys. Rev. Lett.} {\bf 83} (1990 4690.  See also L. Randall and K. Sundrum, {\it Phys. Rev. Lett.} {\bf 83} (1999) 3370 for one brane model (RS1).


\bibitem{fourteen}  P. C. Aichelburg and R. U. Sexl, Gen. Rel and Gravity 2 (1971) 303.

\bibitem{fifteen}  S. B. Giddings, S. Kachru and J. Polchinski, {\it Phys. Rev.} {\bf D66} (2002) 106006.

\bibitem{sixteen}  D. M. Eardley and S. B. Giddings, {\it Phys. Rev.} {\bf D66} (2002) 044011.

\bibitem{seventeen} K. Kang in the 10th International (Blois03) Conference, Helsinki, Finland, June 2003 $(http://www.physics.helsinki. fi/ ~ blois_{-}03/ program.html)$; in ``Small x and Diffraction 2003 Workshop", FNAL, IL, September 2003 (http://conferences.fnal.gov/smallx/index.html). 


\end{thebibliography}
\end{document}